\documentclass{aa}
\newcommand{\be} {\begin{equation}}

\newcommand{\ee} {\end{equation}}

\newcommand{\src}{GRB\,990510}
\newcommand{\VLT}{{VLT--FORS1}}
\newcommand{\N}{{NTT--SUSI2}}
\newcommand{\E}{{ESO 3.6\,m EFOSC2}}
\newcommand{\R}{$R$}
\newcommand{\I}{$I$}
\newcommand{\V}{$V$}

\newcommand{\ot}{OT}
\newcommand{\bc}{\begin{center}}
\newcommand{\ec}{\end{center}}
\newcommand{\rc}{\rm}
\input psfig.tex
\begin{document}

\thesaurus{11.07.1; 11.16.1; 12.03.3; 13.07.1 }

\title{ESO deep observations of the optical afterglow of GRB\,990510 \thanks{Partially 
based on data collected  at the ESO \VLT\ (Paranal), \N\ and \E\ (La Silla) 
telescopes}}

\author{Gian Luca Israel$^{1,}$\thanks{Affiliated to I.C.R.A.}, 
Gianni Marconi$^1$, Stefano Covino$^2$, Davide Lazzati$^{2,3}$, Gabriele 
Ghisellini$^2$, 
Sergio Campana$^{2,\ast\ast}$, Luigi Guzzo$^2$, 
Gianantonio Guerrero$^2$, and Luigi Stella$^{1,\ast\ast}$} 

\institute{
Osservatorio Astronomico di Roma, Via Frascati 33, 
I--00040 Monteporzio Catone (Roma), Italy
\and
Osservatorio Astronomico di Brera, Via E. Bianchi 46, 
I--23807  Merate (Lecco), Italy
\and
Universit\'a degli Studi, via Celoria 16, I--20133 Milano, Italy
}

\date{Submitted 11 June 1999 / Accepted .....}
\offprints{Gianluca.Israel@oar.mporzio.astro.it}
\maketitle
\markboth{Israel et al.}
                  {ESO deep observations of the optical afterglow of \src}

\begin{abstract}
We present the results of optical observations of the \src\ field carried out
at different epochs from European Southern Observatory (ESO) telescopes.   
Deep observations, down to a limiting magnitude 
of about 27 and 24 in the Bessel--\R\ and Gunn--\I\ band, respectively, were  
obtained between May 16 and 18 from the ESO NTT--SUSI2 telescope and on May 20 
from the ESO 3.6\,m (EFOSC2) telescope. 
These observations, together with other published photometric data, 
allowed to monitor the faint tail of the decaying 
Optical Transient (\ot) associated to the \src. 
We discuss the light curves in the different filters (\V, \R\ 
and $I$) in the light of the recently proposed decay laws. 
No obvious host associated to the \src\ optical afterglow was found in 
the \R\ and \I\ band images.  
By comparing the light curves with respect to the theoretical colors of 
galaxies with different morphology we derived a lower limit of $R \sim 26.6$ 
for the host galaxy. 
 
\end{abstract}


\section{Introduction}

On 1999 May 10.36743 UT 
the BATSE detectors on board CGRO, and the Gamma--Ray Burst 
Monitor (GRBM) and the Wide Field Cameras (WFCs) on board the Italian--Dutch 
satellite {\it Beppo}SAX detected a gamma ray burst, \src ,  
with a fluence of 2.5$\times 10^{-5}$ erg cm$^{-2}$ 
above 20 keV (Kippen 1999; Amati et al. 1999; Dadina et al. 1999). 
The first optical follow--up observations began only $\sim$3.5 hours after the 
$\gamma$--ray event and revealed a relatively bright optical 
transient ($R$=$17.54$ , Axelrod, Mould \& 
Schmidt 1999; Vreeswijk et al. 1999a) at  
$\alpha=13^{\rm h}38^{\rm m}07.11^{\rm s}$, $\delta=-80^\circ 
29\arcmin 48.2\arcsec$ (equinox 2000; Hjorth et 
al. 1999b). When compared to previously studied afterglows, 
the \ot\ showed 
initially a fairly slow flux decay ($\propto t^{-0.9}$;  
Galama et al. 1999), that steepened after one day 
($F_\nu\propto t^{-1.3}$; Stanek et al. 1999a) and further 
steepened after 4 days ($F_\nu\propto t^{-1.8}$; 
Pietrzy\'nski \& Udalski 1999; Bloom et al. 1999) and 5 days 
($F_\nu\propto t^{-2.5}$; Marconi et al. 1999a,b).
Such a 
progressive and smooth steepening had not been observed 
before.
Vreeswijk et al. (1999), using the VLT, detected red--shifted 
absorption lines in the \ot\ spectrum corresponding to a 
redshift lower limit of $z=1.619$. 

In this letter we report on deep observations of the \ot\ performed with the
ESO VLT, NTT and 3.6\,m telescopes. 
These observations allowed to extend the coverage of the \ot\ light curve 
up to $\sim$10 days from the burst onset and to search for an underlying 
host galaxy.

\section{Observations and results}

\begin{figure*}[tbh]
\centerline{\psfig{figure=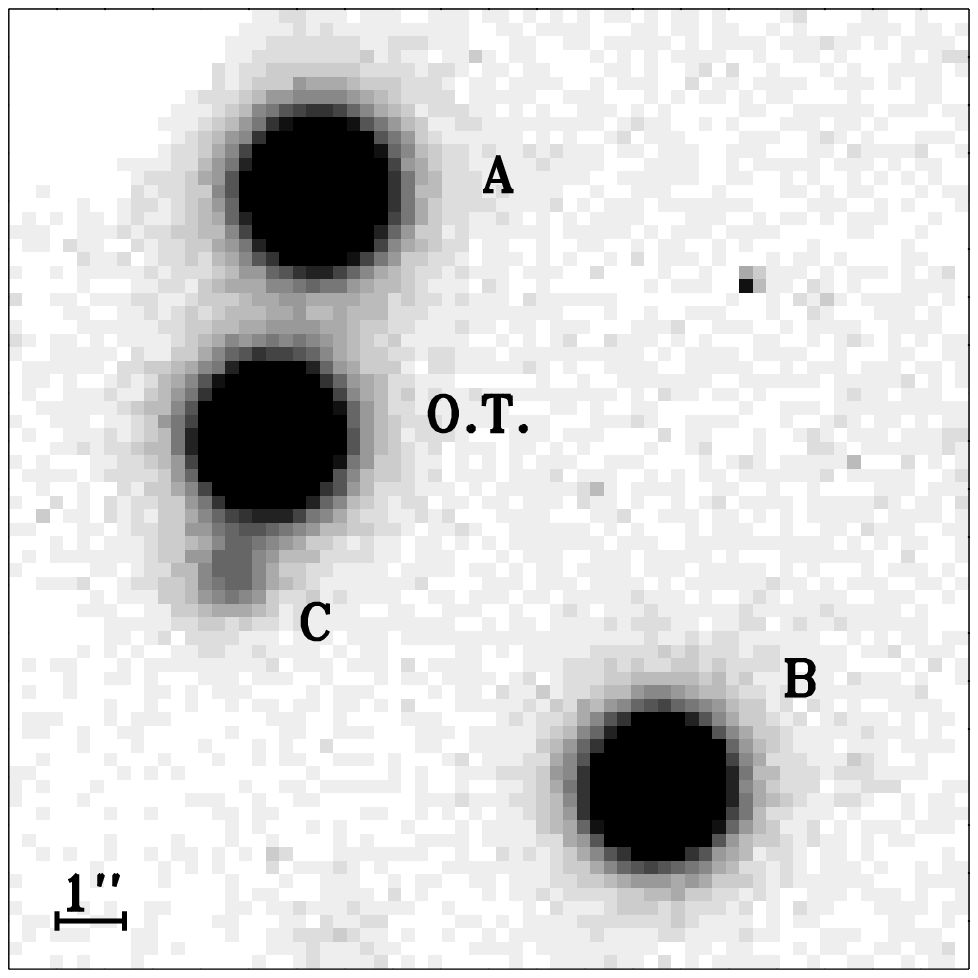,width=8.6cm,height=8.6cm}
\psfig{figure=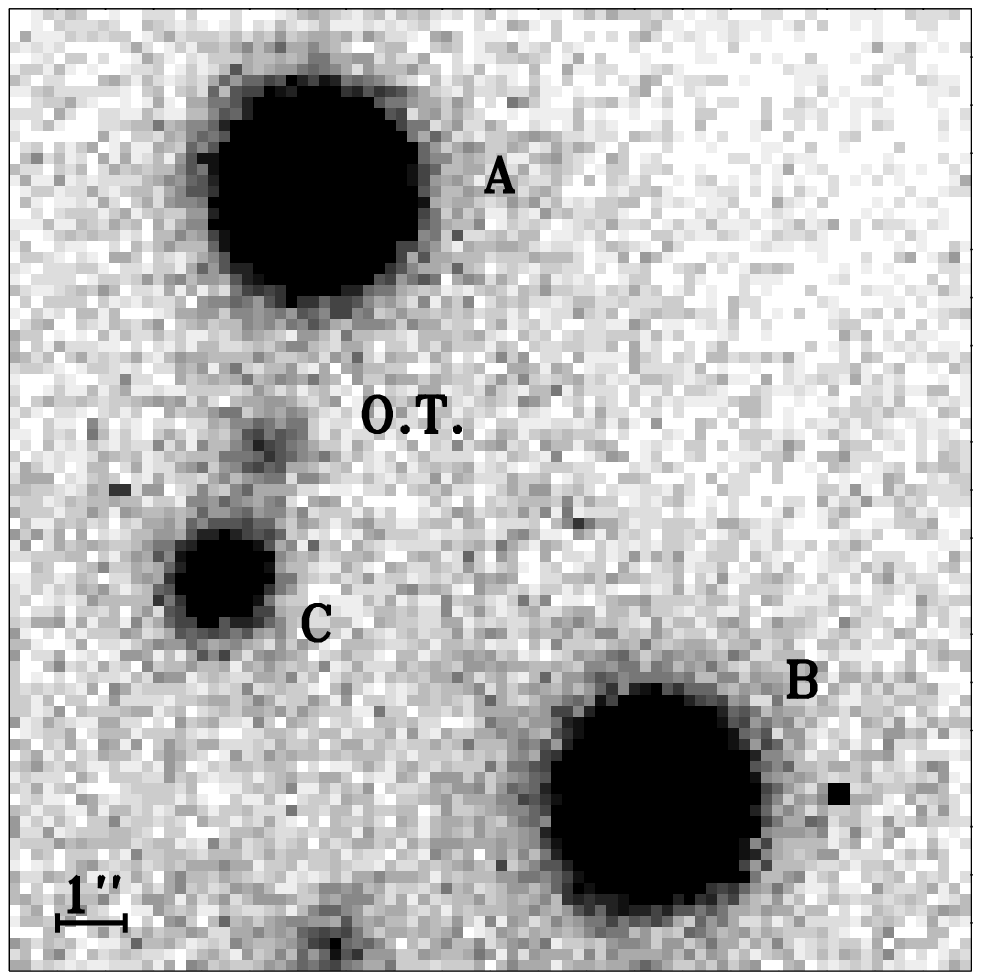,width=8.6cm,height=8.6cm} }
\caption{The field of the \src\ afterglow as seen by the \VLT\ 
in the \R\ band when the \ot\ was $R=19.1$ (left panel; 
600\,s exposure time) and the \N\ in the \R\ band when the \ot\ was $R=23.7$
(right panel; 3000\,s). 
The optical afterglow is marked (\ot).}
\end{figure*}

The observations were performed with the 8\,m VLT--Antu telescope equipped 
with the Focal Reducer/Low Dispersion Spectrograph  (FORS1) on May 
11 (6.8\arcmin$\times$6.8\arcmin\ field of view and  
0.2\arcsec/pixel resolution), the ESO 3.5\,m NTT 
equipped with the Superb Seeing Imager -- 2 (SUSI2) between May 16--18 
(5.5\arcmin$\times$ 5.5\arcmin\ field of view and 0.16\arcsec/pixel 
resolution), and the ESO 3.6\,m  telescope with the ESO Faint Object 
Spectrograph and Camera (EFOSC2) mounted at the F/8 Cassegrain Focus on 
May 20 (5.5\arcmin$\times$5.5\arcmin\ field of view and  
{\rc 0.32}\arcsec/pixel resolution). 
We performed photometry in the Bessel--\R\ and Gunn--\I\ filters. 
The data were reduced using standard {\it ESO--MIDAS} and {\it IRAF} 
procedures for bias subtraction  and flat--field correction. 
Photometry for each stellar object in the image was derived both with the 
DAOPHOT\,II and the ROMAFOT MIDAS--packages (Stetson 1987; 
Buonanno \& Iannicola 1989).  
Point--like source $R$ magnitudes were derived by comparison with
nearby stars, assuming $R=16.5$ for the star at $\alpha = 13^{\rm h}38^{\rm m}
00.82^{\rm s}$, $\delta=-80^\circ 29\arcmin 11.7\arcsec$ {\rc(Bloom et al. 1999), 
while the $I$ magnitudes 
were calibrated observing a number of Landolt photometric standards  
during the observational night (Landolt 1992)}.  
In Fig. 1 the field around the position of the \ot\ as observed by the \VLT\ 
(May 11; left panel) and \N\ (May 18; right panel) telescopes is shown. 
Table~1 reports the results of the photometry for each of the pointings.

\begin{table*}
\begin{center}
\caption{VLT, NTT and ESO 3.6\,m magnitudes of the afterglow.}
\begin{tabular}{lclllcl}
\hline 
Date & &Exposure & Filter& &Seeing&\\
(1999 UT) & Telescope  &(s)&& Magnitude$^a$ & (\arcsec)&Reference \\
\hline
May 11.047 & Antu/\VLT\ & 10  &\R &18.96$\pm$0.02 &1.2&1 \\ 
May 11.136 & Antu/\VLT\ & 600 &\R &19.10$\pm$0.02 &1.2&2\\  
May 16.110 & ESO\N& 1800      &\R &23.0$\pm$0.1   &1.8&3 \\
May 17.107 & ESO\N& 2400      &\R &23.4$\pm$0.1   &1.2&4 \\
May 18.131 & ESO\N& 3000      &\R &23.7$\pm$0.1   &1.1&4 \\ 
\hline
May 20.190 & \E   & 900       &\I &$>$23.6        &1.0&-- \\
\hline \\
\end{tabular}
\end{center}
$^a$ Uncertainties are at 1$\sigma$ confidence level.\\
{\em References:} [1] Covino et al. 1999a; [2] Covino et al. 1999b; [3] 
Marconi et al. 1999a; [4] Marconi et al. 1999b.
\end{table*}

The mediocre seeing of the observations is in part due to the low \ot\ 
elevation at the Paranal and La Silla Observatories.
The May 18 NTT--SUSI2 image is the deepest 
and the one obtained with the best seeing (1.1\arcsec). 
In this image the \ot\ is sufficiently faint to allow for a sensitive search 
for additional underlying point--like or diffuse objects. 
Both the DAOPHOT\,II and the ROMAFOT packages failed to associate a 
point--like Point Spread Function (PSF) to the \ot. 
Moreover the flag that records the sharpness of each single 
object is consistent with diffuse emission either from an object with 
broad--wings or from two point--like nearby objects. 
However the presence of the \ot\ itself plays an important role in 
increasing the background level which, in turn, 
translates into a reduced detection sensitivity. 
An underlying point--like object (either an unresolved host galaxy 
or a faint star) could have been detected if its magnitude were
smaller than $R \sim 26.6$ and with a angular extension $>$ 1\arcsec. 
Note that the HST image of the GRB\,990123 field showed a candidate host galaxy
of $\sim 1\arcsec$ extension (Fruchter et al. 1999).  
If a similar size host were to be expected in our case
(the redshift of the two bursts is similar), this would likely remain 
unresolved in our images.

We also can infer a lower limit for a host galaxy near but PSF--separated from 
the \ot\ by summing the three NTT--SUSI2 images carried out on May 16/17/18:  
a 7200~s exposure image in the \R--band was obtained and analysed. 
Stellar objects in the image were searched with the DAOPHOT\,II; the 
faintest point--like star that we detected have \R$\sim$27 (S/N$\sim$5). 

An alternative way of detecting a host galaxy consists in 
monitoring the low flux end of the afterglow decay: 
the presence of an unresolved galaxy causes the light curves to 
flatten when the flux of the \ot\ becomes comparable to that of the host.
This method was successfully applied to the 
afterglow light curves, obtained 
from days to months after the GRB event, 
in the cases of GRB\,971214 (Odewahn et al. 1998), GRB\,980703 
(Bloom et al. 1998; Castro--Tirado et al. 1999) and GRB\,970508 
(Zharikov \& Sokolov 1999).
Typical values of the known underlying galaxies are in the 
\R=22--27 range (Hogg \& Fruchter 1999 and references therein).

\begin{figure*}[tbh]
\centerline{\psfig{figure=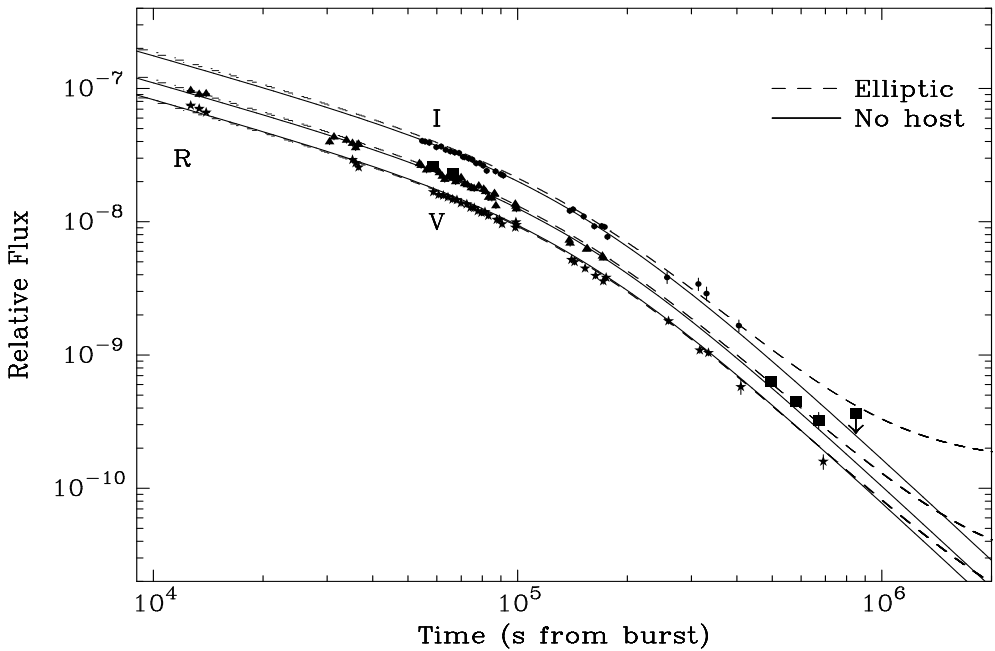,width=15.8cm,height=8.8cm}}
\caption{\src\ afterglow \V, \R\ and \I\ band light curves fitted using 
eq.\,1 under two
different hypothesis: (a) no underlying host galaxy (solid lines), and assuming a $R=26.6$ 
(b) elliptical  galaxy (dashed lines). 
The underlying galaxy is placed at $z=1.6$ and stellar evolution was taken into 
account. Filled squares mark our dataset. See text for details. 
}
\end{figure*}
We applied the same technique to \src. In order to 
further characterise the afterglow decay we collected all the 
published fluxes of the \ot\ in the \V, \R\ and $I$ bands
(Axelrod et al. 1999; Galama et al. 1999; Harrison et al 1999; Kaluzny et al. 
1999a,b; {\rc Stanek 1999}; Stanek et al. 1999a,b; Pietrzy\'nski \& Udalski 
1999a,b,c; Beuermann et al. 1999). {\rc When available we used the known 
uncertainties, while in the remaining cases the 10\% of the flux measurement was 
considered as a typical error value.}
We fitted the $V$, $R$ and $I$ light curves ({\rc 36}, 43 and 32 data points, 
respectively) by using the empirical model 
for the flux evolution, $F_{\nu}(t)$, described in Marconi et al. (1999a) 
\begin{equation}
F_\nu(t)\, =\, 
{k_\nu t^{-\alpha_1} \over 1+(t/t_{\ast})^{\alpha_2-\alpha_1}} \ \ .
\end{equation}
Stanek et al. (1999b; 
hereafter S99) adopted the same model. 
The eariler model proposed by Bloom et al. 
(1999; see also Harrison et al. 1999) is similar. 
$F_{\nu}(t)$ is characterised by four free parameters: two power--law 
indices $\alpha_1$ and $\alpha_2$ (for the earlier and the later time part of 
the decay, respectively), a folding time $t_{\ast}$ where the two power--laws 
match, and the normalisation $k_{\nu}$. 
Table\,2 summarises the results obtained from the fitting. 
Note that parameter uncertainties are 1$\sigma$ 
for a single interesting parameter; all parameters in the fit were left free 
to vary. 
Similar results were reported by S99. We ascribe their somewhat 
smaller uncertainties (and larger $\chi^2$) to 
the fact in evaluating the uncertainties in each parameters S99   
held the other parameters fixed. {\rc We note that a different assumed 
values of the measurement uncertainties may affect the fit and account for  
the different $\chi^2$ values quoted in our work and in S99.}
\begin{table}
\caption{\src\ light curve fit.}
\begin{tabular}{ccccc}
\hline\hline
&&&t$_{\ast}$&\\
Band & $\alpha_1$  & $\alpha_2$ & (days) & $\chi^2/dof^a$ \\
\hline
$R$    & $0.75\pm 0.03$ & ${\rc2.46}\pm 0.06$ & ${\rc 1.3}\pm 0.1$ &1.1 \\
$V$    & ${\rc 0.88\pm 0.03}$ & ${\rc 2.68}\pm 0.13$ & $1.8\pm 0.2$ &0.8 \\
$I$    & $0.77\pm 0.05$ & $2.34\pm 0.11$ & $1.6\pm 0.2$ &0.8 \\
\hline 
\end{tabular}\\
\noindent $^a$ $dof$ stands for degree of freedom.\\
Note --- Uncertainties refer to 1 $\sigma$ confidence level.
\end{table}
Fig.\,2 shows the $V$, $R$ and $I$ {\rc relative flux light curves 
(obtained from M$_{V,R,I}$=--2.5\,log\,F$_{V,R,I}$)}
fitted 
with the model in Eq.~1 keeping fixed 
$\alpha_1$, $\alpha_2$ and $t_{\ast}$ to the best values obtained for 
the $R$ band and leaving free to vary only the normalisation (solid lines). 
A $\chi^2$ of {\rc 95 for 105} degree of freedom ($dof$) corresponding to a 
$\chi^2/dof$ of 0.9  was obtained. 
Note that in this case, given the larger number of data points in the $R$ filter, 
the simultaneous fit in the $V$, $R$ and $I$ bands is biased towards the 
$R$--band light curve. 
The derived fitting parameters are nearly filter--independent,  
within the statistical uncertainties, suggesting a similar  
temporal evolution of the emission in the $VRI$ bands where only the 
normalisation $k_{\nu}$ varies; this was already noted by 
Bloom et al. (1999) and Marconi et al. (1999a).

To include the possibility of an underlying galaxy we added to 
the model described in Eq.~1 a fifth free parameter (i.e. a constant) 
corresponding to the host galaxy flux. 
Galaxy color indices strongly depend on 
the morphology and distance of the host and were adopted from Buzzoni 
(1995, 1999 and references therein) which take into account the effects of stellar 
evolution within the galaxy. 
We consider a set of color index values corresponding to $z=1.6$ and to 
three different morphologies: elliptical (\V--\R=2.4; \R--$I$=2.1), spiral 
(Sb; \V--\R=0.32; \R--$I$=1.02) and irregular (\V--\R=0.15; \R--$I$=0.55). 
Moreover we corrected the set of data for absorption  
in the direction of \src\ ($E_{B-V}\simeq${\rc 0.20; Stanek et al. 1999b}) 
which corresponds to an $E_{V-R}\simeq E_{R-I}\simeq$ 0.15.
 
Dashed lines in Fig.\,2 show the fit obtained by using the colors 
of an elliptical host galaxy of magnitude $R=26.6$. 
We first fitted the light curves as before without the deepest points 
(one in the \V, three in the \R\ and one in the $I$ band), then we added the deep 
data of this paper, keeping fixed $\alpha_1$, $\alpha_2$ and $t_{\ast}$, 
and leaving the normalisation and the host galaxy 
flux parameters free to vary.
In all cases we obtained a $\chi^2/dof$ in the 0.9--1.1 range. 
As expected, the unchanged value of the reduced $\chi^2$ testifies that 
the constant parameter does not significantly  improve the fit.   
In the case of an elliptical host the lower limit on the galaxy 
magnitude is strongly driven by the \ot\ upper limit in the $I$--band; 
any (elliptical) 
object brighter than $R=26.6$ would produce a levelling off of the 
\ot\ $I$--band light curve that is not observed. 
The spiral and irregular host galaxy cases are consistent with the $I$--band 
data, while the lower limit is driven by the VLT deep observation in the 
\V--band (Beuermann et al. 1999); any object brighter than 
$R=26.6$ must be brighter than observed in the \V\ band.
If the host galaxy is farther than $z=1.6$, lower magnitudes are to be  
expected, while color indeces (which are differential quantities) remain 
nearly constant (at least up to $z\simeq2.0$).

\section{Conclusions}

We reported on the deep optical observations of the \src\ \ot. 
We are able to follow the \ot\ in the \R\ band down to 23.7 and derive an 
upper limit in the \I\ band ($>23.6$). 
We propose a functional form for the fitting of the multi-band optical 
light curve (see also S99; Bloom et al. 1999) and derive 
strong upper limits on the magnitude of galaxy hosting the GRB. 
In particular, our photometric data exclude an elliptical host 
galaxy brighter than \R=26.6. {\rc Further observations are needed 
in order to look for a faint R$\geq$27 galaxy host.} 

\begin{acknowledgements} 
The authors thanks G. Iannicola for his kind support with ROMAFOT,  
and the referee K.Z. Stanek, the comment of which helped to 
improve this paper. 
\end{acknowledgements}

\end{document}